\documentclass{article}%
\usepackage{amsfonts}
\usepackage{amsmath}
\usepackage{amssymb}
\usepackage{graphicx}%
\begin{document}

\title{First principles quantum Monte Carlo}
\author{J. M. A. Figueiredo
\and Universidade Federal de Minas Gerais\\
Departamento de F\'{\i}sica CP 702 - Belo Horizonte 30.123-970 -
Brazil}
\date{08/08/2006}

\maketitle

\begin{abstract}
Present quantum Monte Carlo codes use statistical techniques adapted to find
the amplitude of a quantum system or the associated eigenvalues. Thus, they do
not use a true physical random source. It is demonstrated that, in fact,
quantum probability admits a description based on a specific class of random
process at least for the single particle case. Then a first principle Monte
Carlo code that exactly simulates quantum dynamics can be constructed. The
subtle question concerning how to map random choices in amplitude
interferences is explained. Possible advantages of this code in simulating
partial histograms observed in particle diffraction experiments are discussed.

\end{abstract}

\newpage{}

Predictions in quantum mechanics of pure states come from amplitude
calculations. Despite the fact that information concerning particle dynamics
is stored in amplitudes, no experiment can be performed which directly
measures the value of an amplitude. For this reason, probabilities have a
rather unusual source in quantum theory for they come from squared amplitudes
rather than from a random source. In this sense, they must be considered as "a
posteriori" probabilities, since they are obtained after the physical problem
has been solved by using amplitude calculations. In contrast, in classical
physics particle dynamics provided by stochastic processes are calculated by
using "a priori" probabilities, obtained from specific physical models. Thus,
Monte Carlo methods, when applicable to quantum theory, are conceptually
different from those used to simulate classical dynamics. In fact, two classes
of Monte Carlo methods, suitable to meet the requirements of the quantum
theory, were devised. One type uses Monte Carlo integration techniques to
solve numerically the Schr\"{o}edinger equation \cite{Foulkes}. This class
provides the correct wave function for complex many-body problems using some
available mathematical techniques \cite{McMillan},\cite{Kalos},\cite{Ceperley}%
. The other class uses calculated quantum probabilities to simulate particle
motion subject to quasi-classical scattering \cite{Butcher},\cite{Varfo}. In
this case the quantum problem of a basic process is analytically solved and
the probabilities obtained are used to simulate complex situations involving
an ensemble of these basic processes. In both classes, the primary goal is to
obtain some amplitude prior to getting or to using probabilities. Consequently
existing quantum Monte Carlo codes do not simulate quantum dynamics using a
priori probabilities as is usual for real stochastic processes. Both theories
make use of probabilities but they generate completely different Monte Carlo
codes. This reflects the intrinsic reality-free interpretation of quantum
theory which denies any use of sets of classical trajectories (like those
provided by classical Monte Carlo simulations) as a reliable description of
quantum phenomena.

In this sense no devised stochastic dynamics has been able to fully simulate
quantum dynamics. Notwithstanding direct calculation of quantum probabilities
expressed as a sum of histories forming a stochastic-like process was
performed by Marinov \cite{marinov}. He proved that squared amplitudes can be
written as the path integral involving a product of transition matrices
defining histories in a classical-like phase space. Marginal calculations,
obtained from partial integration over momentum variable, lead to histories
which are written as a functional of classical trajectories described by a
Langevin equation. His formalism mimics completely a (classical) stochastic
process, except that resulting transitions matrices are quasi-probabilities
instead a positive-defined transition matrix. The real-valued character of his
transition matrices provides some negative-valued paths that are necessary to
produce quantum interferences. This stochastic-like process cannot be directly
modelled by a Monte Carlo code since the paths are constructed from
quasiprobabilities which cannot be used to draft random numbers, the seed of a
first-principle-defined Monte Carlo code. This is a clear evidence that a
simple stochastic process cannot model any quantum dynamics, as already
demonstrated by Baublitz \cite{baublitz}. In contrast to this immediate
interpretation of Marinov%
%TCIMACRO{\U{b4}}%
%BeginExpansion
\'{}%
%EndExpansion
s results a recent work of Skorobogatov and Svertilov \cite{skorobo} has given
support to the idea of a priori probabilities generating quantum processes.
They demonstrated that quantum probabilities for an isolated two level system
correspond to a particular kind of Chapmann-Kolmogorov equation presenting
both a non-Markovian character and a specific discrete jumping process. The
formal stochastic process they obtained was entirely deduced from quantum
dynamics with no additional hypothesis included. Thus, it appears that is
possible to mimic some types of quantum problems by a suitable class of formal
stochastic process. These results rise some relevant questions that ask for a
solution. It is necessary to solve the apparently contradictory results of
Baublitz and Skorobogatov and Svertilov, in order to understand which class of
quantum problems admits a formal stochastic representation and to determine
whether is possible or not to write a quantum Monte Carlo code based on a
priori probabilities. Positive answers to these questions would probably lead
to new possibilities on computation techniques of quantum problems and, at a
more fundamental level, could give some clues about the still unsolved
questions concerning the subquantum world.

In this work, we touch upon and partially answer these questions. We present a
specific class of formal stochastic process whose time evolution forms a chain
mediated by quasiprobability matrices. Then it is demonstrated that Marinov
functional is a particular instance of these matrices. In sequence, a priori
transition probabilities that generate these real-valued transition matrices
are constructed and an associated Monte Carlo code can then be written using
them. Differently from usual quantum Monte Carlo calculations the obtained
code is fully based on a classical algorithm without the need of make
reference to amplitude calculations although still capable of fully simulate
quantum dynamics. In the code described here, quantum probabilities are
obtained as an excess probability above a proper level, to be defined below.
The delicate question of how to explain amplitude interference in terms of a
proper choice of random numbers is discussed and its possibility demonstrated
for the first time. Since a Monte Carlo code is the ultimate and definitive
description of a random process, the existence of such a code, capable of
model quantum processes, puts in order the need of a stochastic justification
to quantum theory.

This work meets the requirements on presenting a first principle Monte Carlo
code based on a formal stochastic dynamics. It has, in this sense, the
operational goal of solve problems in quantum mechanics, but also pursues the
challenge of to give clues to important questions involving the very nature of
the subquantum world. In what follows we develop a model of specific
stochastic processes presenting dynamics mediated by real-valued transition
matrices. Then we present arguments that proves how a particular form of
Marinov%
%TCIMACRO{\U{b4}}%
%BeginExpansion
\'{}%
%EndExpansion
s results match this model. As a result a positive-defined transition matrix
is derived from quantum theory and a complete chain Monte Carlo process is
then trivially obtained.

A Monte Carlo code works upon choices of random numbers, forming histories. A
process is simulated by the use of specific rules defining the way choices are
selected as valid and on the way histograms are calculated. Thus the bare
result of a Monte Carlo code is a set of valid counts. The difficulty in
treating quantum events using a Monte Carlo code comes from the possibility of
interference effects. For consider two quantum states labeled $A$ and $B$,
with amplitudes $\Psi\left(  A\right)  $ and $\Psi\left(  B\right)  $ and let
$\rho=\left\vert \Psi\left(  A\right)  +\Psi\left(  B\right)  \right\vert
^{2}$ the total probability for the state $\Omega=\Psi\left(  A\right)
+\Psi\left(  B\right)  $. This probability can be null even if the probability
for individual states, $\left\vert \Psi\left(  A\right)  \right\vert ^{2}$ and
$\left\vert \Psi\left(  B\right)  \right\vert ^{2}$, is not null. This means
that we cannot take the number of counts generated by event described by state
$A$ alone and separately sum to the number of counts generated by event $B$ in
order to get the histogram for the total probability.

In order to write a code adapted to quantum mechanics it is necessary to know
how to handle, generically, process of type $0=s+\left(  -s\right)  $, where
$s$ is a number of counts in a histogram. Since there is no "negative"
histories it is immediate that no code, based on direct choice of random
numbers, may generate this result. In what follows we introduce the main
argument of this work. It proves that, in an extended probability space (EPS),
a code providing this result in fact exist. Let%
%TCIMACRO{\U{b4}}%
%BeginExpansion
\'{}%
%EndExpansion
s consider the case $s=kw$, where both $k,w\in\left[  -1,1\right]  $. Choose
numbers $g\in\left[  0,2\right]  $ and $\lambda,v\in\left[  -1,1\right]  $
such that $k=1-g$ and $w=\lambda-v$. Define $K=\left(  1-g\right)  I$, where
$I$ is the two-dimensional identity matrix and write a set of vectors, $W,Po$
and $V$, related by
\begin{align*}
Po  &  \equiv%
\begin{pmatrix}
\lambda\\
1-\lambda
\end{pmatrix}
;V\equiv%
\begin{pmatrix}
v\\
1-v
\end{pmatrix}
\\
W  &  \equiv%
\begin{pmatrix}
w\\
-w
\end{pmatrix}
=%
\begin{pmatrix}
\lambda-v\\
\left(  1-\lambda\right)  -\left(  1-v\right)
\end{pmatrix}
=Po-V,
\end{align*}
The product giving $s$ is equal to $\left(  KW\right)  _{0}$. Since the
elements in $Po$ and $V$ are positive and sum to one they can be treated as
probabilities. In similar way we define a quasiprobability vector $S$ from the
number $s$ and get, from vector $W$, the result:
\begin{align*}
S  &  \equiv%
\begin{pmatrix}
s\\
-s
\end{pmatrix}
=KW=KPo-V+g%
\begin{pmatrix}
v & v\\
1-v & 1-v
\end{pmatrix}
Po\\
&  \Rightarrow P\equiv S+V=%
\begin{pmatrix}
1-g\left(  1-v\right)  & gv\\
g\left(  1-v\right)  & 1-gv
\end{pmatrix}
Po
\end{align*}
The columns in the last matrix sum to one. Their elements are positive if
$g\left(  1-v\right)  \leq1$ and $gv\leq1$. Both relations demand that
$g\leq2$, a condition naturally fulfilled. For $K\geq0$ $\left(
g\leq1\right)  $ any value of $\nu\in\left[  0,1\right]  $ is allowed. For
$K<0\left(  1<g\leq2\right)  $ allowed values of $v$ are restricted to the
interval $\left(  \left(  1-g\right)  g^{-1},g^{-1}\right]  $. Within these
restrictions, a positive matrix $M$ is obtained that satisfies $P=MPo$.
Therefore equation $P=MPo$ maps directly into a formal stochastic process
describing a system possessing two states subject to random transitions
between them \cite{hoel}. In this process the system in the state indexed as
$1$ and possessing probability $Po_{1}$ may transit, with probability $gv$, to
the state indexed as $0$; if state is $0$ the system may transit, with
probability $g\left(  1-v\right)  $, to state $1$. Total probability for state
$0$ after this random process is: $P_{0}=$ $\left(  1-g\left(  1-v\right)
\right)  Po_{0}+gvPo_{1}=\left(  MPo\right)  _{0}$. A first principle Monte
Carlo code, using these transition probabilities and standard programming
technics, may model vector $P$. Now run $H$ histories of this code. The number
of them ending on state $0$ is $P_{0}H=\left(  s+v\right)  H$ and on state $1$
is $P_{1}H=\left(  1-v-s\right)  H$. Assume $s>0$. In discarding the first
$vH$ histories ending in state $0$ results in the value $Hs$. This is a
permitted operation in a Monte Carlo code for it is one type of criterion used
for validation of a history. Thus the number $s>0$ can be properly simulated
using a first principle Monte Carlo code. If $s<0$ its absolute value is found
from similar reasoning applied to state $1$, from which $\left(  1-\nu\right)
H$ histories are discarded.

We now treat the case $0=s+\left(  -s\right)  $. Modelling the event
$s+\left(  -s\right)  =kw+k(-w)$ means that a random choice must select event
$s$ or event $-s$ and run a history for the chosen process. The resulting
histogram is just addition of the obtained histograms. Select the same value
$v$ for both processes and consider a set of $H$ histories. Select event $s$
or $-s$ with equal probability. The histogram modelling event $s$ equals to
$\left(  S+V\right)  H/2$ and event $-s$ has a histogram equal to $\left(
-S+V\right)  H/2$. Total histogram, which describes the occurrence of both
events, equals to $\left(  S-S+2V\right)  H/2=VH$. Thus, in discarding the
first $\left(  VH\right)  _{0}$ histories results in exactly null counts of
the histogram to state $0$. That is, this code can model the algebraic sum
$0=s+\left(  -s\right)  $, in the sense that well-defined algorithms are
associated individually to these events and its joint occurrence. Using a
proper criterion for validation of histories, a result of null counts is
generated. we now show how to use this method in a form appropriate to model
real quantum problems.

The starting point is the time-slice Feynman formula
\[
\Psi\left(  x,t\right)  =\left(  \frac{m}{2\pi i\hbar\epsilon}\right)
^{\frac{n}{2}}\int\exp\left(  \frac{i}{\hbar}\mathcal{A}_{n}\right)  \Psi
_{0}\left(  x_{0}\right)  \prod\limits_{l=0}^{n-1}dx_{l}%
\]
which is know to converge to the Schr\"{o}edinger wave function when
$t/n=\epsilon\rightarrow0$ \cite{gosson}. In this equation the symbol
$\mathcal{A}_{n}$ is the discrete-time action written as
\[
\mathcal{A}_{n}=\epsilon\sum_{l=0}^{n-1}\left[  \frac{m}{2\epsilon^{2}}\left(
x_{l+1}-x_{l}\right)  ^{2}-U\left(  x_{l}\right)  \right]  \equiv
\mathcal{A}_{n}\left\{  x_{l}\right\}
\]
with end point $x_{n}=x$. The point chosen to calculate the potential rises
subtle questions \cite{gosson} and the simplest version, but not necessarily
the more precise one, is assumed here. The squared amplitude, written in
non-dimensional units $x\rightarrow x\sqrt{\hbar\epsilon/m}$; $U\rightarrow
\hbar/\epsilon U$, becomes%
\begin{equation}
\left\vert \Psi\left(  x\right)  _{n}\right\vert ^{2}=\left(  \frac{1}{2\pi
}\right)  ^{n}\int\limits_{-\infty}^{\infty}\exp\left(  i\left[
\mathcal{A}_{n}\left\{  x_{l}\right\}  -\mathcal{A}_{n}\left\{  x_{l}^{\prime
}\right\}  \right]  \right)  \Psi_{0}\left(  x_{0}\right)  \Psi_{0}^{\ast
}\left(  x_{0}^{\prime}\right)  \prod\limits_{l=0}^{n-1}dx_{l}dx_{l}^{\prime}
\label{Psi2}%
\end{equation}
As before it is convenient to recast this equation by setting $x_{l}%
=u_{l}+.5w_{l}$ and $x_{l}^{\prime}=u_{l}-.5w_{l}$. Expansion of the argument
of the exponential on these new variables puts the action in a new form:
instead the two-time chain present in the kinetic energy term, a three-time
chain on the variable $u$ appears in association with a single-time dependence
on $w$. The result is a odd series in this variable of the form
\begin{gather*}
\mathcal{A}_{n}\left\{  u_{l}+\frac{1}{2}w_{l}\right\}  -\mathcal{A}%
_{n}\left\{  u_{l}-\frac{1}{2}w_{l}\right\}  =\\
-\sum_{l=1}^{n-1}\left[  \left(  u_{l+1}+u_{l-1}-2u_{l}\right)  w_{l}+U\left(
u_{l}+\frac{1}{2}w_{l}\right)  -U\left(  u_{l}-\frac{1}{2}w_{l}\right)
\right]  -\\
\left(  u_{1}-u_{0}\right)  w_{0}-U\left(  u_{0}+\frac{1}{2}w_{0}\right)
+U\left(  u_{0}-\frac{1}{2}w_{0}\right)
\end{gather*}
Partial integration, for indices greater than zero and at independent time
slices, of the exponential that appears in eqn$\left(  \ref{Psi2}\right)  $ is
admissible. Due to the odd parity in $w$, it results in a real-valued kernel
for the probability, valid for individual time slices, and given by%
\begin{multline}
K_{l}\left(  u_{l+1}+u_{l-1}-2u_{l}+\frac{d}{du_{l}}U\left(  u_{l}\right)
;u_{l}\right)  =\nonumber\\
\frac{2}{\pi}\int\limits_{0}^{\infty}\cos\left(  \left(  u_{l+1}%
+u_{l-1}-2u_{l}+\frac{d}{du_{l}}U\left(  u_{l}\right)  \right)  w_{l}%
+2\sum_{k=1}^{\infty}\frac{w_{l}^{2k+1}}{\left(  2k+1\right)  !}\frac
{d^{2k+1}U\left(  u_{l}\right)  }{du_{l}^{2k+1}}\right)  dw_{l} \label{kernel}%
\end{multline}
In the above equation we have a kernel valid for short time intervals. Time
slice zero deserves a special issue. The first coefficient of $w_{0}$ is just
the velocity written in terms of variables $u$. Since there is no chain term
in $w_{0}$, integration in this variable can be performed separately and has
the form%
\[
\frac{1}{2\pi}\int\limits_{-\infty}^{\infty}e^{i\left(  u_{1}-u_{0}\right)
w}\Psi\left(  u_{0}+\frac{1}{2}w\right)  e^{-iV\left(  u_{0}+\frac{1}%
{2}w\right)  }\Psi^{\ast}\left(  u_{0}+\frac{1}{2}w\right)  e^{iV\left(
u_{0}-\frac{1}{2}w\right)  }dw=W\left(  u_{0},u_{1}-u_{0}\right)
\]
This leads to a modified Wigner function at time zero: an additional phase
term changes the value of the initial wave function. Its existence does not
make invalid the line of reasoning presented here so we postpone discussion on
its meaning. Putting together the last results it is clear that quantum
probability can be written as a path integral of real valued kernels
possessing as initial condition a Wigner function . Explicitly we have%
\begin{equation}
\left\vert \Psi\left(  x\right)  _{n}\right\vert ^{2}dx=dx\int\limits_{-\infty
}^{\infty}\prod\limits_{l=2}^{n-1}K\left(  u_{l+1}+u_{l-1}-2u_{l}+\frac
{d}{du_{l}}U\left(  u_{l}\right)  ;u_{l}\right)  W\left(  u_{0},u_{1}\right)
\prod\limits_{l=0}^{n-1}du_{l} \label{Marinov}%
\end{equation}
which is a particular case of a more general result obtained by Marinov
\cite{marinov} who get similar expression but using integrals in the phase
space instead. We now approximate these integrals by a discrete sum on a
finite spatial grid. In this case each path consists of a product of type
$\left[  \Delta x\prod\limits_{l=2}^{n-1}K_{l}\Delta u_{l}\right]  W\Delta
u_{0}\Delta u_{1}\equiv k_{ef}\tilde{W}$, where $k_{ef}=\Delta x\prod
\limits_{l=2}^{n-1}K_{l}\Delta u_{l}$ and $\tilde{W}=W\Delta u_{0}\Delta
u_{1}$. The sum over all paths becomes a sum over all values of the product
$k_{ef}\tilde{W}$. Both are real numbers with absolute values smaller than
one. Thus the formalism of extend probability space just described is
applicable to every path entering in the above sum. A Monte Carlo code that
selects a path by chance and simulates the product $k_{ef}\tilde{W}$ in the
extended space generates a count that approximates to the result of
eqn$\left(  \ref{Marinov}\right)  .$

In fact more information is available by eqn$\left(  \ref{kernel}\right)  $
defining the kernels. Since the kernel $k_{ef}$ is factorable, the first
product in a path has the form $K_{1}W=\left[  M_{1}\left(  P_{0}-V\right)
\right]  _{0}=\left(  P_{1}-V\right)  _{0}$ where the last equality comes from
the fact that the reference vector $V$ (of a given path) is invariant under
application of the matrix $M$. The second product has the form $K_{2}%
K_{1}W=K_{2}\left(  M_{1}P_{0}-V\right)  _{0}=\left(  M_{2}M_{1}%
P_{0}-V\right)  _{0}$. Recursive application of this method on the product
involving the whole chain of a path leads to%
\begin{equation}
\left\vert \Psi\left(  x\right)  _{n}\right\vert ^{2}dx\approx\left(
P_{n}-V_{path}\right)  _{0}=\left[  \prod\limits_{\substack{l=2\\path}%
}^{n-1}M_{l}\left(  P_{0}-V_{path}\right)  \right]  _{0} \label{Qprob}%
\end{equation}
This result clearly express that probabilities in quantum theory are, in the
EPS formalism, a difference of probabilities. Using the invariance property of
$V$, we get $P_{n}=\prod\limits_{path}M_{l}P_{0}$. This equation represents a
classical probability process in the EPS. Quantum effects comes from two
effects: a renormalization of the probability and from the swapping of states
in the EPS. Notice the argument in the transition matrices $M\left(
u_{l+1}+u_{l-1}-2u_{l}+\frac{d}{du_{l}}U\left(  u_{l}\right)  ,u_{l}\right)  $
is a time-difference equation%
\[
u_{l+1}+u_{l-1}-2u_{l}+\frac{d}{du_{l}}U\left(  u_{l}\right)  =y_{l}%
\]
It is possible to interpret this expression as a Langevin equation possessing
noise source $y$ (with uniform probability distribution) and linking the
various time slices, in such a way that the transition matrix is function of a
stochastic force through the random variable $y$. This way, total probability
at time $t$ is given by%
\[
P\left(  x\right)  =\sum_{\left\{  path\right\}  }\prod\limits_{l}M\left(
y_{l};u_{l}\right)  P_{0}%
\]
where all paths must finish at $x$. Thus, in the EPS formalism, a particle in
quantum motion follows a simultaneous two-step process. It is subject to a
classical random process displayed in the Langevin equation. The same random
variable that drives the stochastic force performs the swapping process in the
EPS. Initial probability is calculated from the addition of a proper reference
value to the Wigner function that makes the result everywhere positive. The
associate Monte Carlo method, with $H$ histories, has the following structure:

\begin{itemize}
\item a spatiotemporal grid is mounted and approximate Marinov kernels calculated;

\item from the knowledge of these kernels an appropriate reference value is
determined for each path;

\item transition matrices are calculated;

\item initial joint probability $P_{0}$ is constructed from the Wigner function;

\item initial positions at time zero and one are selected from the initial
joint probability $P_{0}$;

\item a random number is selected;

\item it determines the next position, calculated form the Langevin equation;

\item in the same time step swapping of states are calculated using the
probability values displayed in the matrix $M\left(  y;u\right)  $;

\item sequential choices of random numbers lead to a path, all them restricted
to terminate at position $x$;

\item normalized histogram for the states, at each ending point, is generated
from the counts;

\item quantum probability is get from the excess probability above the
appropriate reference level; one state always has probability above its
reference level and the other one below;

\item another end-point is chosen.
\end{itemize}

Thus, as anticipated, a Monte Carlo code capable of calculate quantum
probabilities and using first principle transition matrices can be
constructed. In addition, the general form of the short-time Feynmam
propagator led to a naive interpretation for the associated quantum
probability path integral, by which space dynamics (conveniently described by
a Langevin equation) matches the swapping process involving the states of the
EPS, both process driven by one single random variable.

Due to the dependence of the eqn$\left(  \ref{kernel}\right)  $ on the value
of the potential energy at time zero, the exact definition of the initial
quasiprobability rises the still elaborate question on (quantum) state
preparation \cite{ballentine}, since this procedure also depends on the way a
potential changes the phase of the initial wave function. This technical
question affects the exact value of the zero-time amplitude but does not
introduce conceptual changes in the algorithm proposed here. Another important
feature is the non-local character of the expression for the quantum
transition quasiprobability, the kernel $K$, expressed by the presence of
high-order derivatives in its argument:%
\begin{equation}
K\left(  y,u\right)  =\frac{1}{\pi}\int\limits_{-\infty}^{\infty}\cos\left(
yw+2\sum_{k=1}^{\infty}\frac{w^{2k+1}}{\left(  2k+1\right)  !}\frac
{d^{2k+1}U\left(  u\right)  }{du^{2k+1}}\right)  dw \label{Mkernel}%
\end{equation}
and this imply that information concerning the existence of a potential is
spread over the whole space even for very well localized potential profiles.
Then it might be possible to capture some of the non-local effects existing in
quantum theory as in the very relevant case of EPR-Bell experiments.
Furthermore note that eqn$\left(  \ref{Mkernel}\right)  $ is valid even for
time-dependent potentials which just adds a index to the potential energy
calculated in the spatiotemporal grid but does not introduces additional
chaining effects. Thus the equivalence of the code proposed here and single
particle quantum dynamics in one dimension is complete for a wide class of
problems, including time-dependent potentials and arbitrary initial
conditions, possibly with inclusion of technical complications induced by
state preparation.

The scenario is clear. Likewise the explanation given for a stochastic case,
in the present interpretation of the results of Marinov, quantum mechanics
resembles a ensemble of systems existing in a state of the EPS at values
consistent with the reference level (the "vacuum"). Measurements makes sense
only for difference of probabilities, relative to vacuum values, as shown in
eqn$\left(  \ref{Qprob}\right)  $. Here, the interpretation of quantum
probabilities gets more involved than the usual one. Quantum probability is a
positive-definite quasiprobability which describes the likelihood of a
specific internal state (the observable one), at some place. This means, in
the present interpretation, that quantum theory cannot give complete account
of the actual physical state of a particle. In fact, the exchange of internal
states during the time evolution of the stochastic dynamics leads to the
effect of interferences of quasiprobabilities provided by the amplitude
formalism. Only one internal component of the stochastic process is observable
because the other has probabilities necessarily below the associated vacuum
level, thus always presenting a negative quasiprobability, so the particle
appear to present no internal state at all. This effect hidden the stochastic
nature of the process, when described by using amplitude calculations, because
the action of individual transition probabilities no longer exists in the
diagonal evolution of the quasiprobability vectors. In consequence the
formalism of quantum theory have less information about the actual dynamics of
the particle than the stochastic one. It may be argued that no measurements
can detect this effect, but as will shown bellow, for a specific class of
experiments, it seems that the stochastic model may capture more details than
the amplitude formalism. This way we have here a clear and precise description
of the importance of trajectories in quantum theory, a rather involved
question on the very foundations of quantum mechanics. When simulated by a
first principle Monte Carlo code, quantum trajectories appear to be real in
the same sense they are, for instance, in codes modelling diffusion. All paths
are tested and included in the statistics. All paths starts from a initially
distributed region of the space and must get the same final observation point.
In this aspect we have a complete equivalency to classical codes, describing
by usual stochastic dynamics. The vacuum level, which determines measurements
results, matters in the quantum case and has no meaning in classical dynamics.
Besides, much more trajectories are used to get the probability at a point,
namely those involving the extra (unobservable) degree of freedom. These two
elements, which are correctly considered in the logical structure of the
extended probability space theory, make quantum trajectories to appear as
possessing no physical reality. That%
%TCIMACRO{\U{b4}}%
%BeginExpansion
\'{}%
%EndExpansion
s the essence of quantum interferences, in the present formalism.

As far as the arguments presented here go, we have demonstrated that
non-relativistic quantum mechanics of one particle admits the existence of a
formal Monte Carlo code based on a priori probabilities. In common to other
Monte Carlo applications, this code have full advantage over analytical
methods when complex mechanical problems are treated. No physical
justification was given to the random choice based on extra degrees of freedom
nor to the mathematical properties of transition probabilities either. Despite
this limitation, it permits a very general modeling of quantum phenomena, and
represents a new class of quantum Monte Carlo method. It may be especially
useful in modelling single hits experiments like that performed by Tonomura,
Endo, Matsuda and Kawasaki \cite{tonomura}. They shot single electrons past a
double slit apparatus and observed the formation of the interfering pattern.
In this case squared amplitudes give only the final histogram valid for a
large number of hits. The code proposed here, when extended to the two
dimensional case, can explain classes of intermediate histograms, valid for
arbitrary number of detected particles, indicating that, in this case, more
information is get using a stochastic process than using amplitude calculations.

The nature of the noise source deserves a special issue. This question
extrapolates the model developed, which is consistent with information
contained uniquely inside the amplitude formalism. Its physical reality relies
on the strong sensitivity of Marinov quasiprobability on the potential energy
as well as the possible existence of extra degree of freedom, both of which
demands proper justification on experimental grounds. Another point touches
the way simulated trajectories are interpreted. It is well known that
trajectories in quantum mechanics are continuous, as shown by Feynman
\cite{feymman} and the same does happen in ordinary Brownian motion, this fact
granted by the Lindeberg condition \cite{gardiner}. Despite these similarities
quantum dynamics cannot be explainable by any stochastic force associated to
simple classical noise sources unless some nonclassical procedure are used
such as a negative diffusion coefficient \cite{nelson}. The extra degrees of
freedom introduced here and vacuum renormalization of the stochastic
probability, which are acceptable logical elements inside a classical
reasoning, appear to complement the properties of pure classical noise sources
in order to reproduce quantum effects. This way we have got a consistent
conciliation of both models and explained the origin of its widely different
behavior as well the common probabilistic nature, namely the dynamics provided
by the Langevin equation. This argument effectively solves the apparently
contradictory results presented in ref. \cite{baublitz} and \cite{skorobo}.
Since it is now possible to interpret quantum theory through a proper
stochastic view the possibility of a real noise source driving quantum
phenomena cannot be discarded at all. Such possibility demands additional
studies in more complex situations, as those involving the Dirac equation or
many-particle systems, in order to explore more deeply the possible existence
of a hidden stochastic mechanism in these cases as well. Probably, additional
properties of the transition matrices may appear when these cases are
considered giving new formal clues concerning the nature of the vacuum source
underlying quantum phenomena.

\end{document}